# ON THE EFFECT OF ULTRAVIOLET IRRADIATION OF ALANINE, GLYCINE AND TRYPTOPHAN CRYSTALS ON RAMAN SPECTRA OF THESE MATERIALS

Kompan M.E., Malyshkin V.G.,
A.F.Ioffe Institute, St. Petersburg

V.V. Chechel
Academic University by J.I. Alferov, St. Petersburg

O.Yu. Tsybin
St. Petersburg Polytechnic University of Peter the Great

**Abstract** The effect of ultraviolet radiation (275 nm) on the Raman spectra of amino acid microcrystals: alanine, glycine, tryptophan was studied. The position of some lines in the spectra after irradiation has slightly shifted towards high energies. The main effect: observation of changes in the relative intensities of the components of the Raman spectra after irradiation of microcrystals. It has been suggested that this change is due to the reorientation of some molecules under the influence of radiation.

**Introduction**

Amino acids are relatively simple biomolecules that are part of many other more complex biomolecules, such as proteins, DNA. Molecules of some amino acids were found in outer space on asteroids [1], which is partly an argument in favor of the panspermia hypothesis [2]. The question of the possibility and conditions of the origin of a life as such has been discussed since time immemorial (now we will formulate this as a problem of the emergence of the simplest biomolecules). In the 20th century, the attempts were even made to obtain experimental evidence of the possibility of the origin of organic substances from inorganic [3]. However, the possibility of the existence of something and the participation of this in further processes depends not only on the probability of formation, but also on the probability of the opposite process - on the rate of degradation, decomposition. This seems to be especially relevant in relation to the conditions for the location of molecules on asteroids, in outer space. This work is a step towards addressing this issue.

Amino acids are important not only in the context of global problems. Practical medicine recognizes the important role of amino acid balance in maintaining the homeostasis of the human body. Preparations containing amino acids such as glycine, tryptophan are used as corrective bioactive additives. In this case, the question of the stability of drugs and amino acids in them goes into a specific practical plane. Works in this direction are known [4]. Among other things, there are publications on the creation of facilities for targeted research in this direction [5]. There are also attempts to study the effect of ultraviolet radiation on inorganic materials [6].

Of course, the number of amino acids themselves, their states - in solution, in a crystal, in the form of adsorbed molecules on the surface, or in a living organism, as well as many possible degradation factors are large, and there is no way to consider much in one study. This paper examines the effect of ultraviolet irradiation of microcrystals of alanine, glycine and tryptophan on Raman spectra of these materials

**Experimental technique and samples under investigation.**

Raman scattering spectra were recorded using a HORIBA-JOBIN-IVON MRS 360 micro-Raman instrument. Scattering was excited by the He-Ne light of the 6328.1 Å laser and recorded in the backscattering geometry. The excitatory and recording lights were not polarized. The intensity of the excitation light before the last lens was 0.5-1 mW.

Microcrystals of amino acids with a typical size of several tenths of a millimeter were isolated for manual examination under a microscope from a dried drop of an appropriate solution on a glass substrate. The separated microcrystals were then fixed to a substrate with a plastic adhesive.

The microcrystals were exposed to UV light emitted at 275 nm. The flux density was controlled by a surface barrier photodiode FDUK-2 and was 3-5 mW/cm2. The duration of irradiation was 240 minutes.

**Experimental results**

No fundamental changes in the Raman spectra were recorded in our experiments. To detect relatively weak effects, the procedure was as follows: the initial spectrum of the sample was recorded. After the irradiation session, the spectrum was recorded again. Reliable comparison of the spectra was hindered by the fact that in some cases a non-monotonic background appeared in the spectra of irradiated samples (see, for example, Fig. 5) and, due to this, the assessment of line intensities could be quite subjective. For this reason, the authors did not use a numerical characterization of the change in line intensity and limited themselves to emphasizing qualitative differences.

Each figure in the article shows two spectra - before and after irradiation (red and blue, respectively), either for the same sample, or averaged over the same group of microcrystals. The above spectra had their background subtracted and in some cases scaling was used to facilitate visual comparison. It should be noted that in most cases, a non-linear structureless background was observed in the spectra of irradiated samples, which was noticeably stronger than the background in the spectra of the original samples. This fact is not reflected in the presented spectra.

Fig. 1 shows the spectrum of scattering by alanine microcrystal in the energy region of 70-450 cm$^{-1}$. It can be seen that the band of about 105 cm$^{-1}$ has undergone significant changes. Initially the band consisted of two components and the more energetic band was predominant in the spectrum for the original microcrystal. The spectrum of the crystal after irradiation demonstrate a fairly well resolved doublet.

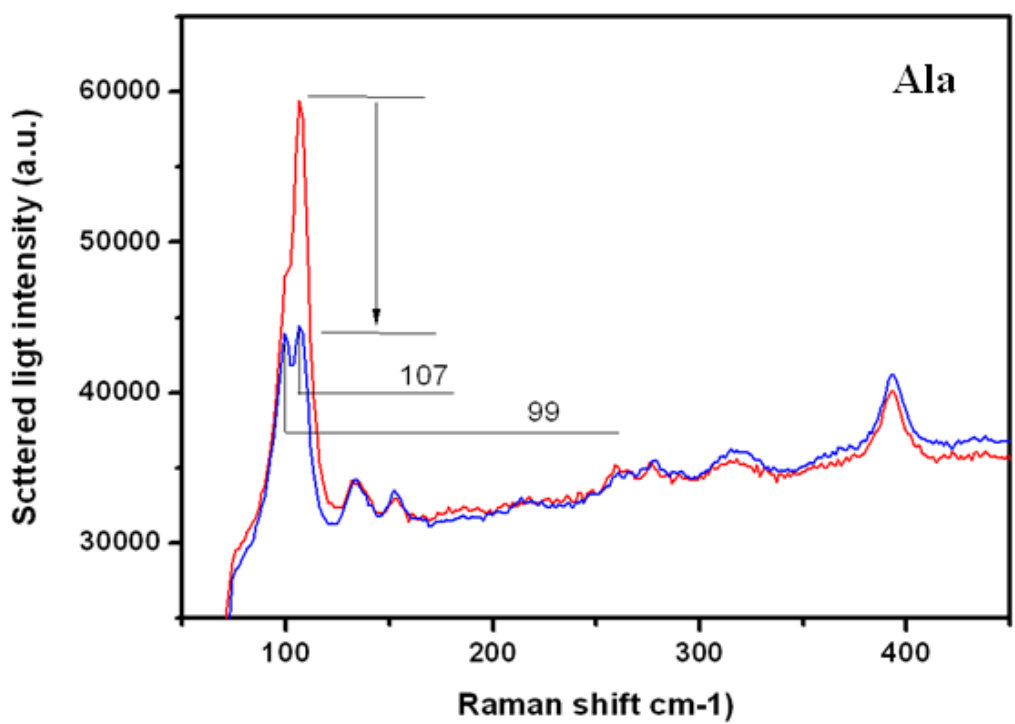

*Fig.1 Raman spectra by alanine microcrystal in the region of 70-450* cm$^{-1}$. *Red curve - before irradiation, blue - after.*

Figure 2 shows Raman scattering spectra of alanine (averaged over a group of microcrystals) in the region of 1100-1700 cm$^{-1}$ / It can be seen that the relative intensity of some lines does not change, while the relative intensity of lines 1297 cm-1 and 1477 cm-1 has decreased, and the intensity of line 1455 cm-1 has increased relative to other lines of the spectrum. When considering spectra on a more detailed scale, a small shift to higher energies after irradiation of the former two lines (of the order of 1 cm$^{-1}$) is noticeable, but line 1455 cm$^{-1}$, on the contrary, had shifted to smaller energy.

In the area 2900-3000 cm$^{-1}$ . the difference between the spectra before and after irradiation is also distinguishable without special treatment. It should be emphasized that the changes in the relative intensity of the components in the spectra of two separate microcrystals in these cases are opposite.

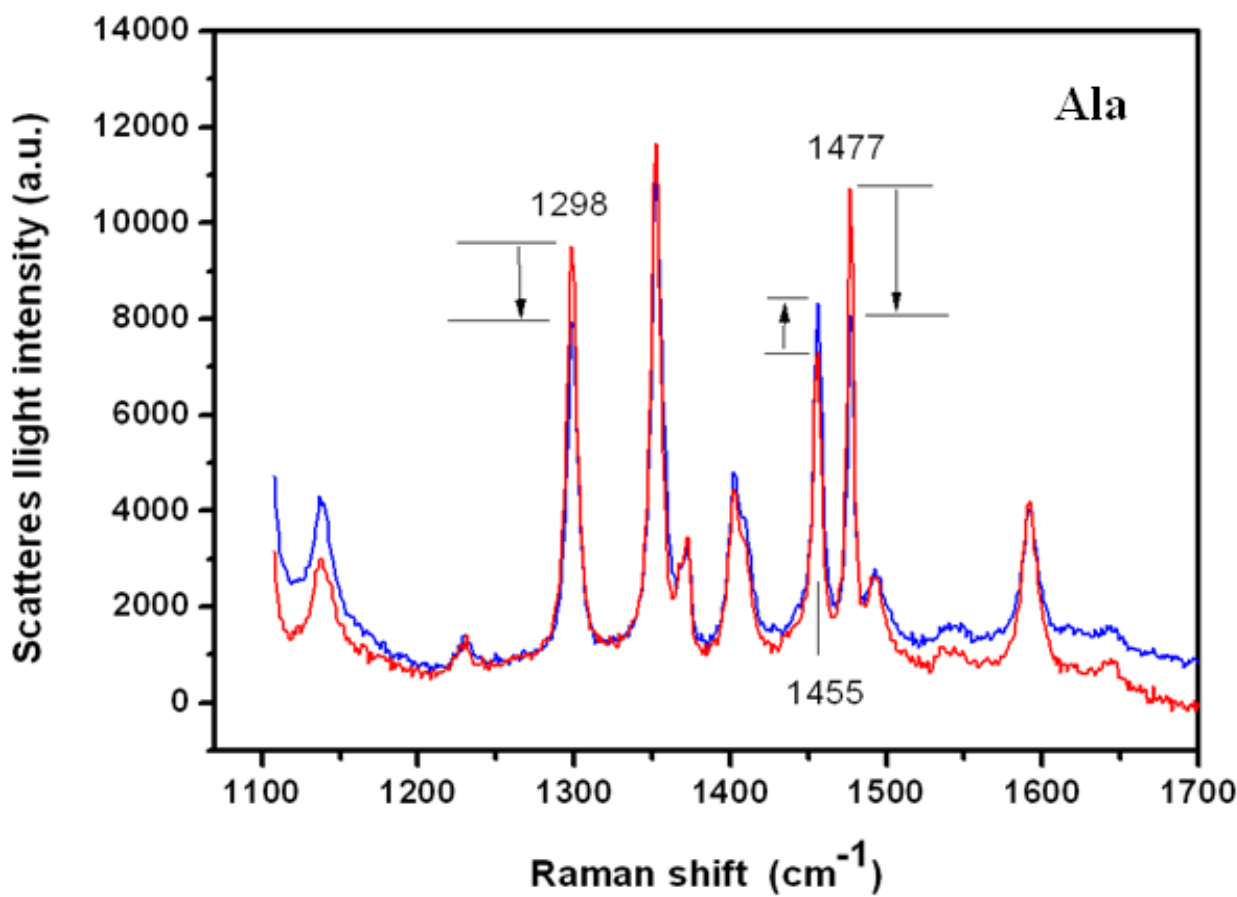

*Fig. 2 Raman scattering spectrum of alanine microcrystals in the region of 1100-1700 arr. see red curve - before irradiation, blue - after. The spectrum is averaged over several microcrystals.*

Similar effects - changes in the relative intensity of the components were recorded on the scattering spectra by microcrystals of other amino acids.

Figure 4 shows the spectra of scattering by glycine microcrystal. Here, too, the relative intensities of the lines have changed as a result of irradiation. The intensities of the central group of lines changed relatively little, while the intensities of lines 103.5 cm$^{-1}$ and 889 cm$^{-1}$ against this background increased by tens of percent.

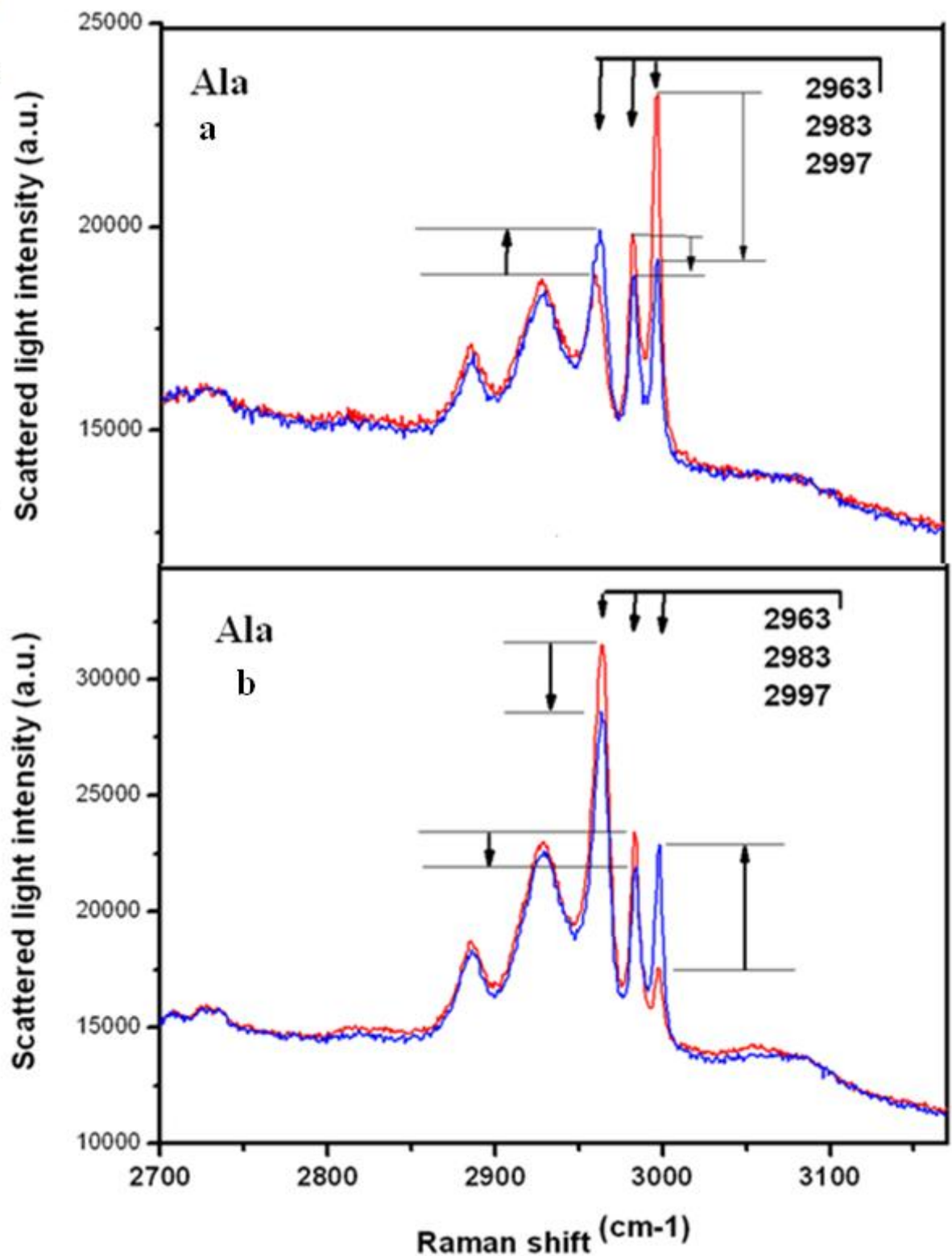

*Fig. 3 a, b. Spectra of individual alanine microcrystals in the region of 2700-3200 cm$^{-1}$. before and after irradiation. Figure a) and Figure b) show spectra for different microcrystals*

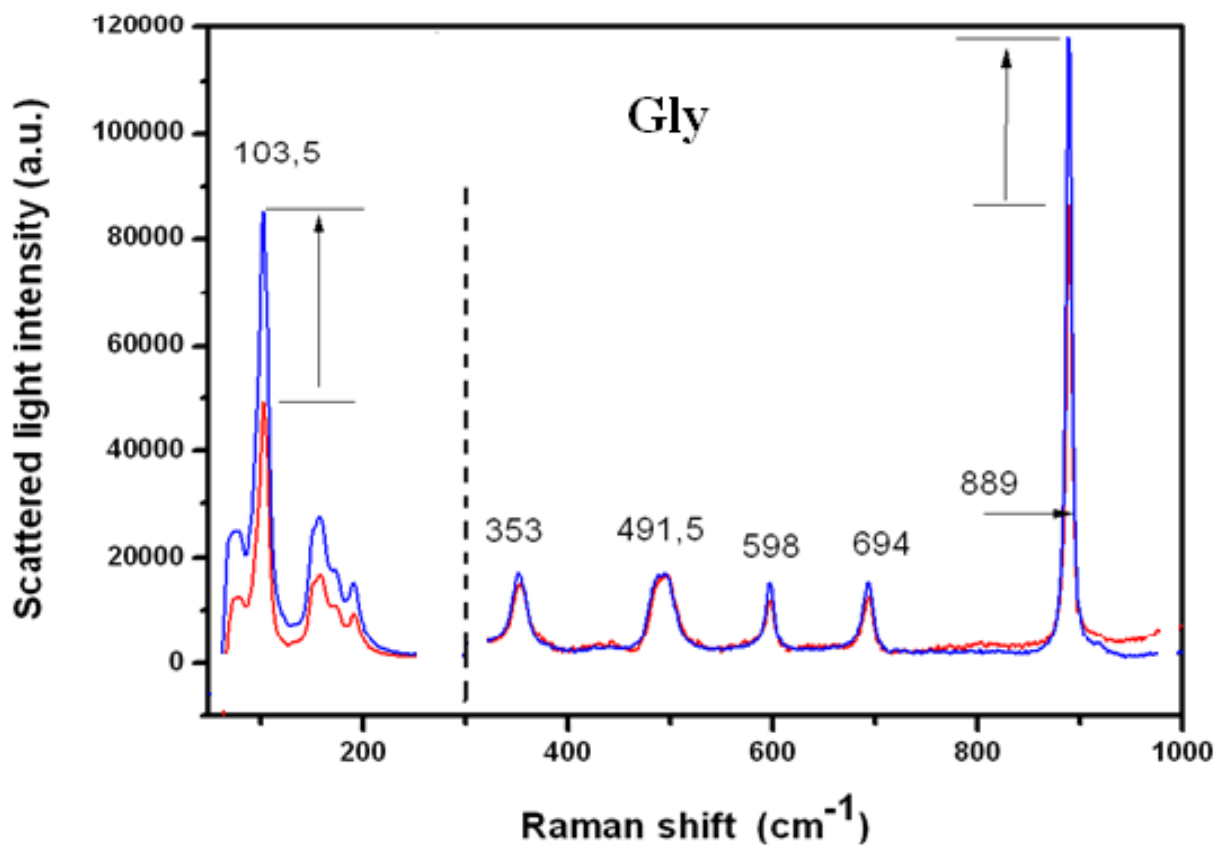

*Fig. 4 Scattering spectra of glycine microcrystals in the region of 70-1000 cm$^{-1}$. The intensity of the lines on the left side of the figure is reduced by 4 times*

In the following figure 5a, c. shows the scattering spectra of tryptophan microcrystals. The upper figure compares spectra the original and irradiated crystal. At the bottom - comparison of the spectrum of the unirradiated crystal and the scattering spectrum from the tryptophan microcrystal obtained from the degraded solution (yellowed) from time to time without additional exposure.

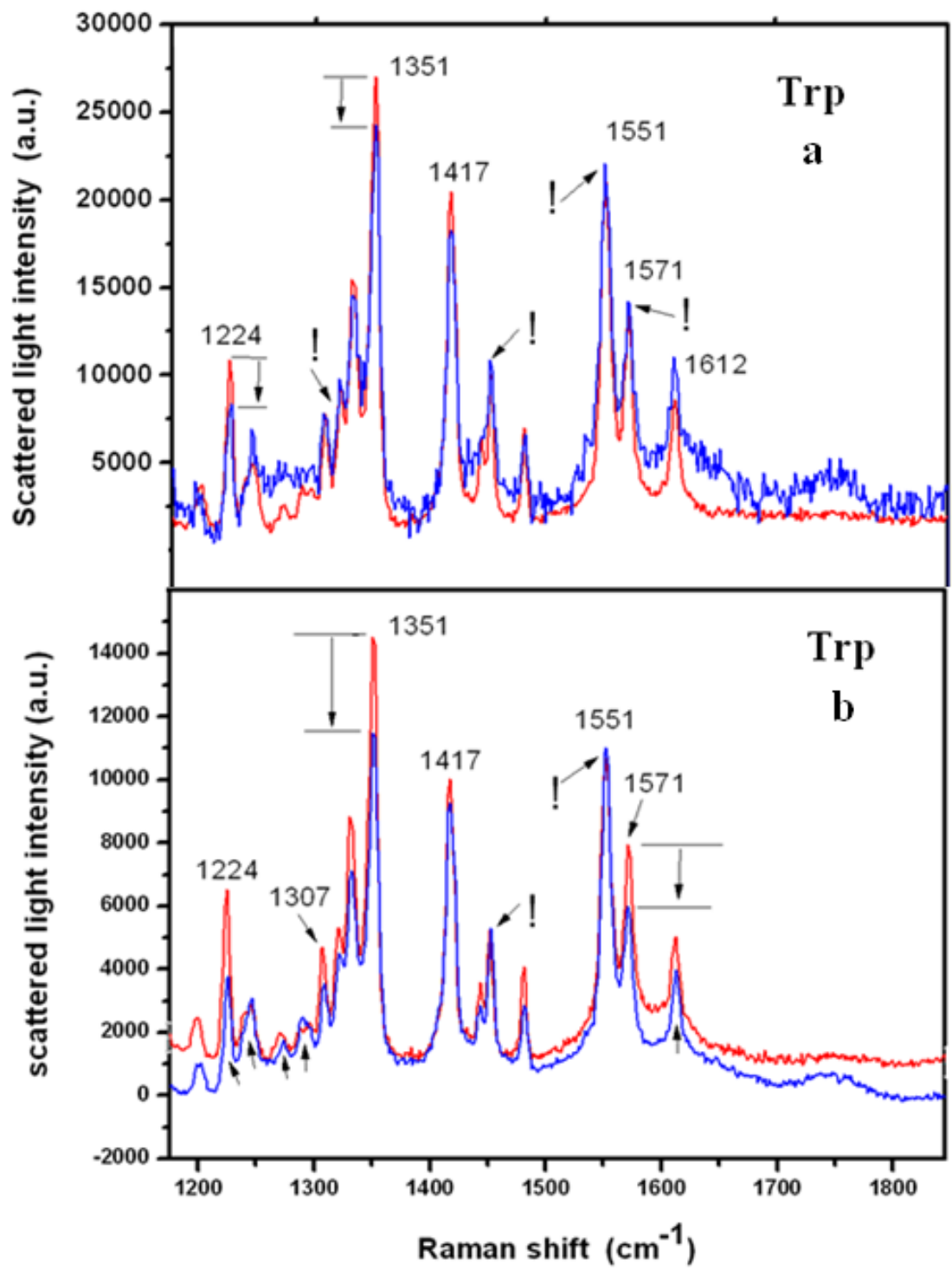

*Fig.5 a, b Raman scattering spectra by tryptophan microcrystals. Fig. a (upper) comparison of the spectrum of the initial crystal and the spectrum obtained after irradiation. Fig. b - comparison of the spectra of the microcrystal of the starting material and the microcrystal of tryptophan from a naturally aged solution. Exclamation marks are placed on lines for which the relative intensity has not changed.*

It can be seen that the relative intensity of the main part of the lines in Fig. 5 decreased after irradiation, while the relative intensities of lines 1551, 1571 cm-1 and some others, highlighted in the figure with exclamation marks, remained unchanged. In fig. 5b, the spectrum of the same initial crystal is compared with the spectrum of a microcrystal obtained by drying a solution of tryptophan that has turned yellow after long-term storage. It can be seen that the changes in the relative intensities of the lines in these two cases are similar.

**Discussion of results.** The change in the Raman spectra under the influence of UV can be caused by a number of reasons. In [5], the possibility of photoinduced chemical reactions is considered as possible mechanisms. In the case of photolysis with the separation of molecules into smaller fragments, one can expect disappearance of lines responsible for a certain type of bonds. In more complex cases, as in [7], a change in the spectrum of tissues *in vivo* may indicate the synthesis of other molecules. It should be taken in mind that the spectra may also depend on changes in the environment of the molecules under study.

In general, from the results of the work, it should be noted that the microcrystals of the studied amino acids are resistant to ultraviolet radiation in relatively small doses.

The ultraviolet wavelength used falls into the absorption band of most organic molecules, so that excited states under the influence of radiation should be formed. The main effect is a change in the relative intensity of the lines, which can be associated with reorientations of molecules in crystals. Such reorientation can follow different mechanisms. By absorbing a quantum of light and then returning to the ground state, the molecule may be in a different orientation, which may be facilitated by the existence of several modifications of crystalline structures for one type of amino acid.

As far as can be concluded from the data of the experiments, no decomposition of molecules and the formation of foreign substances were found. The reason for this is an unchanged set of spectrum components with almost unchanged positions.

The slight changes in positions observed in some cases are most likely due to the heterogeneous broadening of the lines and the predominant reorientation of molecules at certain positions. Perhaps the poorly structured background under the lines in Fig. 5a in the spectrum of the irradiated tryptophan microcrystal indicates the beginning of the material decomposition process.

**References**

.

1. J.T. Wickramamasinghe, C. Wickramamasinghe, W.M.Napier «Comets and the origin of life »//World Scientific, 2010, ISBN-13 978-981-256-635-5, 232 p.

.2. Horneck, Gerda; Klaus, David M.; Mancinelli, Rocco L. (2010). «Space Microbiology». Microbiology and Molecular Biology Reviews. 74 (1): 121–156. doi:10.1128/MMBR.00016-09.

.3. A. I. Oparin "Origin of Life "//1924 (1959) Military Publishing House, p. 38) (Rus: А.И.Опарин «Происхождение жизни»)

.4. A. Zhirasek, H.G. Shulze, C.H. Gughesman, A.L. Sgrig, C.A. Haines, M.V. Blades, R.F.B. Thurner "Discrimination between U Vradiation-induced and termallyinduced spectral changes in AT-paired DNA oligomers using resonance raman spectroscopy "//J. Raman Spectroscopy 2006, 37, 1368-1380. DOI: 10.1002 ./jrs 1552.

.5. V.G.Borio, A.U.Fernandes, L.Silveira Jr (2015) «Characterization of an ultraviolet irradiation chamber to monitor molecular photodegradation by Raman spectreoscopy »// Instrumentation science and technology.

.6. H.Jia, G.Chen, W.Wang «UV radiation-induced Raman spectra changes in lead silicate glasses »//Optical Materials 2006, 29, 445-448. DOI:10.1016/j.optmat.2005.09,078

.7. M.G. Tosato, D.E. Orallo, S.M. Ali, M.S. Churio, A.A. Martin, L. Dikelio "Confocal raman spectroscopy *in vivo* biochemical changes in the human skin by topical formation under UV Radiation "//Photochemistry and Photobiology (B), 2015, 8, 30 pp

# О ВЛИЯНИИ УЛЬТРАФИОЛЕТОВОГО ОБЛУЧЕНИЯ КРИСТАЛЛОВ АЛАНИНА, ГЛИЦИНА И ТРИПТОФАНА НА СПЕКТРЫ КОМБИНАЦИОННОГО РАССЕЯНИЯ ЭТИХ МАТЕРИАЛОВ

*Компан М.Е., Малышкин В.Г.,*
*Физико-технический институт им. А.Ф.Иоффе, Санкт-Петербург*

*Чечель В.В.*
*Академический университет им. Ж.И.Алферова, Санкт-Петербург*

*Цыбин О.Ю.*
*Санкт-Петербургский Политехнический университет Петра Великого*

**Аннотация**

Исследовано влияние ультрафиолетового облучения (275 нм) на спектры комбинационного рассеяния микрокристаллов аминокислот: аланина, глицина, триптофана. Положение некоторых линий в спектрах после облучения незначительно сместилось в сторону больших энергий. Основной эффект: обнаружение изменения относительных интенсивностей компонент рамановских спектров после облучения микрокристаллов. Высказано предположение, что это изменение обусловлено переориентацией части молекул под действием облучения.

**Введение**

Аминокислоты – относительно простые биомолекулы, входящие составной частью во многие другие более сложные биомолекулы, такие как белки, ДНК. Молекулы некоторых аминокислот были обнаружены в космическом пространстве на астероидах [1], что отчасти является доводом в пользу гипотезы панспермии [2]. Вопрос о возможности и условиях зарождения жизни как таковой дискутировался с незапамятных времен (сейчас мы будем формулировать это как проблема возникновения простейших биомолекул). В 20-м веке даже были предприняты попытки получения экспериментальных доказательств возможности зарождения органических веществ из неорганических [3]. Однако возможность существования чего-либо и участия этого в дальнейших процессах зависит не только от вероятности образования, но и от вероятности обратного процесса - от скорости деградации, разложения. Особо это представляется актуальным по отношению к условиям нахождения молекул на астероидах, в космическом пространстве. Настоящая работа является шагом к рассмотрению данной проблемы.

Аминокислоты важны не только в контексте глобальных проблем. Практическая медицина осознает важную роль баланса аминокислот в поддержании гомеостаза человеческого организма. Препараты, содержащие такие аминокислоты как глицин, триптофан используются как корректирующие биоактивные добавки. В этом случае вопрос стабильности препаратов и аминокислот в них переходит в конкретную практическую плоскость. Работы в этом направлении известны [4]. В том числе имеются публикации по созданию установок для целенаправленных исследований в этом направлении [5]. Также известны работы по исследованию влияния ультрафиолетового облучения на неорганические материалы [6].

Безусловно, количество самих аминокислот, их состояний – в растворе, в кристалле, в виде адсорбированных молекул на поверхности, или в живом организме , а также множество возможных факторов деградации - велико, и нет возможности рассмотреть многое в одном исследовании. В данной работе исследуется влияние ультрафиолетового облучения микрокристаллов аланина, глицина и триптофана на спектры комбинационного рассеяния этих материалов

**Экспериментальная техника и исследовавшиеся образцы.**

Спектры рамановского рассеяния регистрировались с помощью установки для микрорамановских исследований HORIBA-JOBIN-IVON MRS

360. Рассеяние возбуждалось светом He-Ne лазера с длиной волны 6328,1 Å и регистрировалось в геометрии обратного рассеяния. Возбуждающий и регистрирующий свет не были поляризованы. Интенсивность света возбуждения перед последним объективом составляла 0,5-1 мВт.

Микрокристаллы аминокислот с типичным размером в несколько десятых миллиметра выделялись для исследований вручную под микроскопом из высохшей на стеклянной подложке капли соответствующего раствора. Затем отделенные микрокристаллы пластичным клеем фиксировались на подложке.

Микрокристаллы подвергались действию ультрафиолетового излучения светодиода, излучавшего на длине волны 275 нм. Плотность потока излучения контролировалась поверхностно-барьерным фотодиодом ФДУК-2 и составляла 3-5 мВт/см$^2$. Длительность облучения составляла 240 минут.

**Результаты экспериментов**

Кардинальных изменений рамановских спектров в наших экспериментах зарегистрировано не было. Для выделения относительно слабых эффектов процедура состояла в следующем: регистрировался исходный спектр образца. После сеанса облучения спектр регистрировался снова. Надежному сравнению спектров препятствовало то, что в некоторых случаях в спектрах облученных образцов появлялся немонотонный фон (см. например, рис.5) на фоне которого оценка интенсивностей линий могла быть достаточно субъективной. По этой причине авторы не использовали численную характеристику изменения интенсивности линий и ограничивались подчеркиванием качественных различий.

На каждом рисунке в статье приведены два близких спектра – до и после облучения (красный и синий соответственно), либо для одного и того же образца, либо усредненные по одной и той же группе микрокристаллов. У приведенных спектров был вычтен фон и в некоторых случаях использовано масштабирование для удобства визуального сравнения. Следует отметить, что в большинстве случаев в спектрах облученных образцов наблюдался нелинейный бесструктурный фон, который был заметно сильнее фона в спектрах исходных образцов. Этот факт не отражен в представленных спектрах.

На рис.1 представлен спектр рассеяния монокристаллом аланина в области энергий 70-450 см$^{-1}$. Видно, что полоса около 105 см$^{-1}$ претерпела значительные изменения. Можно видеть, что исходно полоса состояла из двух компонент, и более энергичная полоса была преобладающей в спектре для исходного микрокристалла. Спектр микрокристалла после облучения представляет собой достаточно хорошо разрешенный дублет.

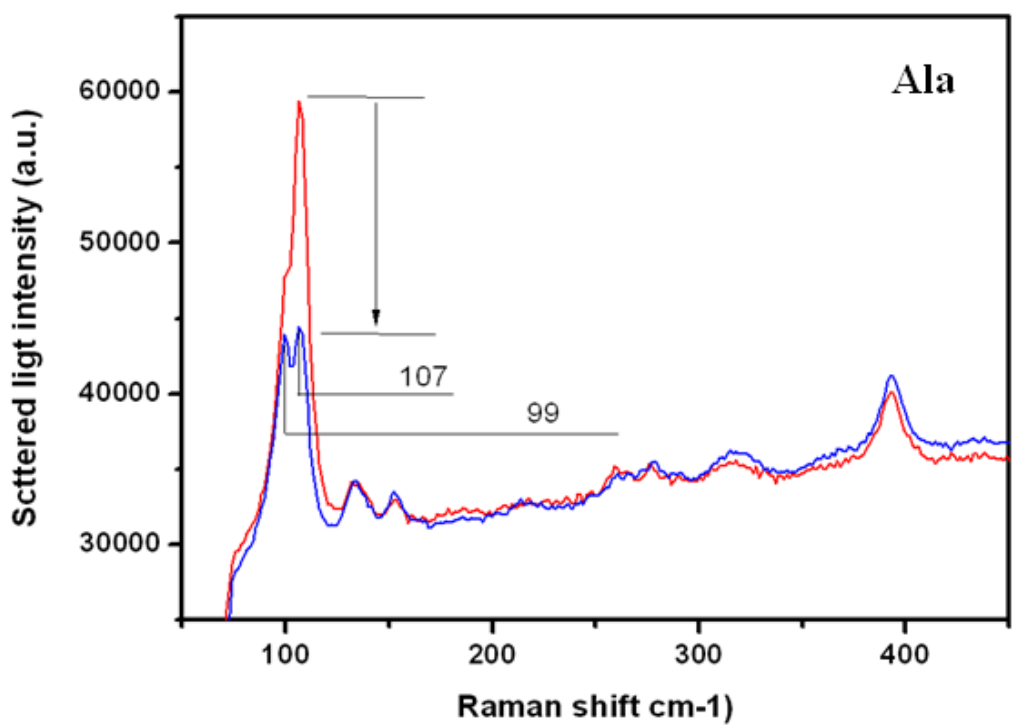

*Рис.1 Спектры рамановского рассеяния микрокристаллом аланина в области 70-450 обр. см. Красная кривая – до облучения, синяя – после.*

На рис.2 представлены спектры рамановского рассеяния того же аланина (усредненный по группе микрокристаллов) в области 1100-1700 см$^{-1}$. Наглядно видно, что относительная интенсивность некоторых линий не меняется, в то время как относительная интенсивность линий 1297 см$^{-1}$ и 1477 см$^{-1}$ уменьшилась, а интенсивность линии 1455 см$^{-1}$ возросла относительно других линий спектра. При рассмотрении спектров в более детальном масштабе заметен небольшой сдвиг (порядка 1 см$^{-1}$) после облучения двух первых линий в бо́льшие энергии, а линии 1455 см$^{-1}$ наоборот, в меньшие.

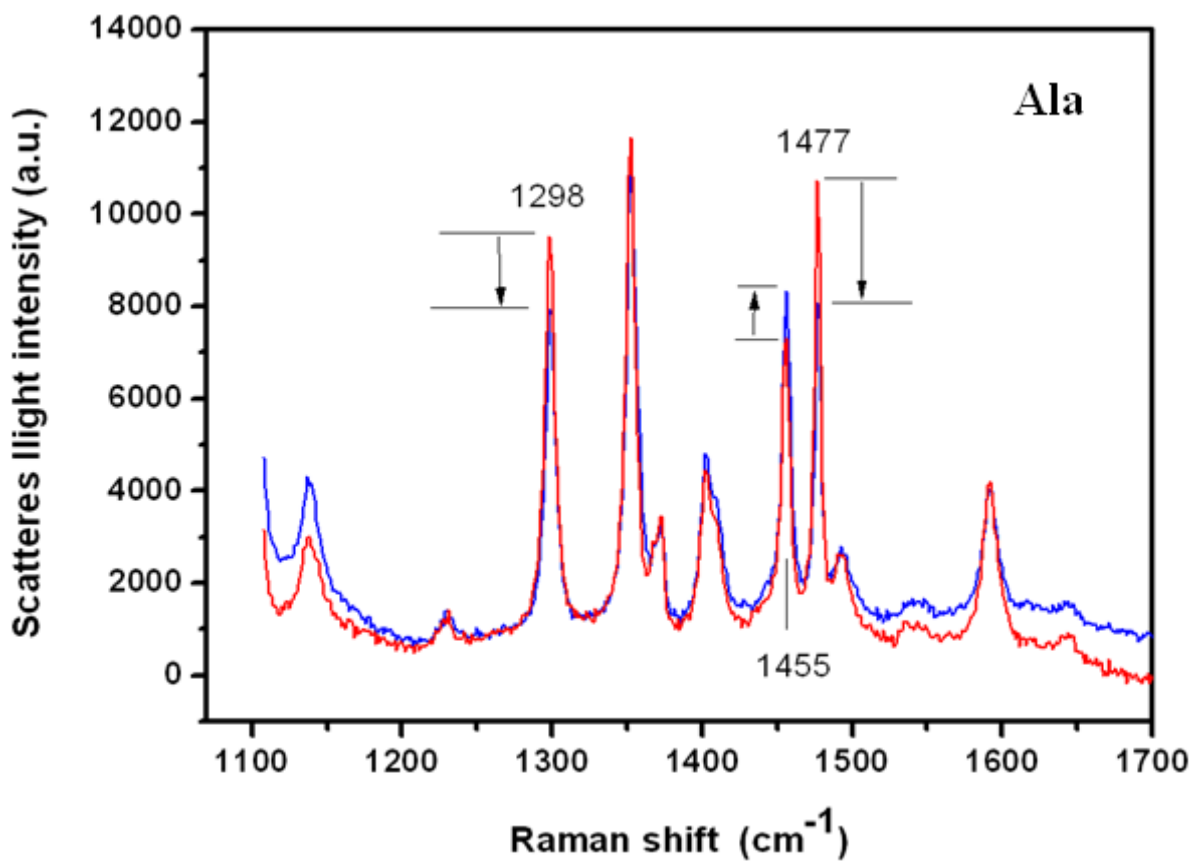

*Рис.2 Спектр рамановского рассеяния микрокристаллов аланина в области 1100-1700 обр. см. Красная кривая – до облучения, синяя – после. Спектр усреднен по нескольким микрокристаллам.*

В области полос 2900-300 обр.см. отличие спектров до и после облучения также различимо без специальной обработки. Необходимо подчеркнуть, что изменения относительной интенсивности компонент в спектрах двух отдельных микрокристаллов этих случаях имеют противоположный характер.

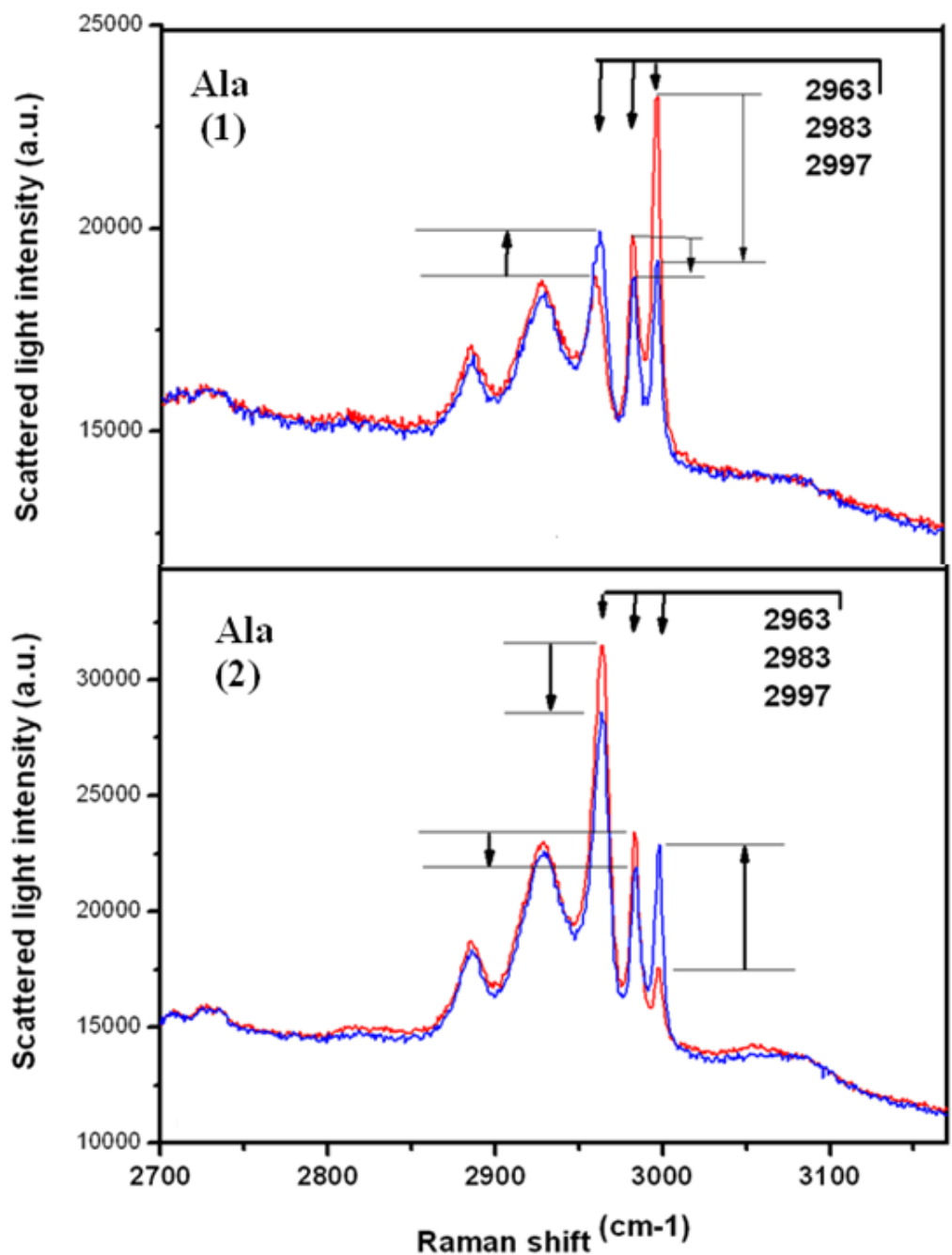

Рис.3 а,в . Спектры отдельных микрокристаллов аланина в области 2700-3200 обр.см. до и после облучения . На рис. а) и рис.в) результаты для разных микрокристаллы

Аналогичные эффекты – изменения относительной интенсивности компонент были зафиксированы на спектрах рассеяния микрокристаллами других аминокислот.

На рис.4 показан спектр рассеяния микрокристаллом глицина. Здесь также в результате облучения изменились относительные интенсивности линий. Относительно мало изменились интенсивности центральной группы линий, в то время как интенсивности линий 103,5 см$^{-1}$ и 889 см$^{-1}$ на этом фоне выросли на десятки процентов.

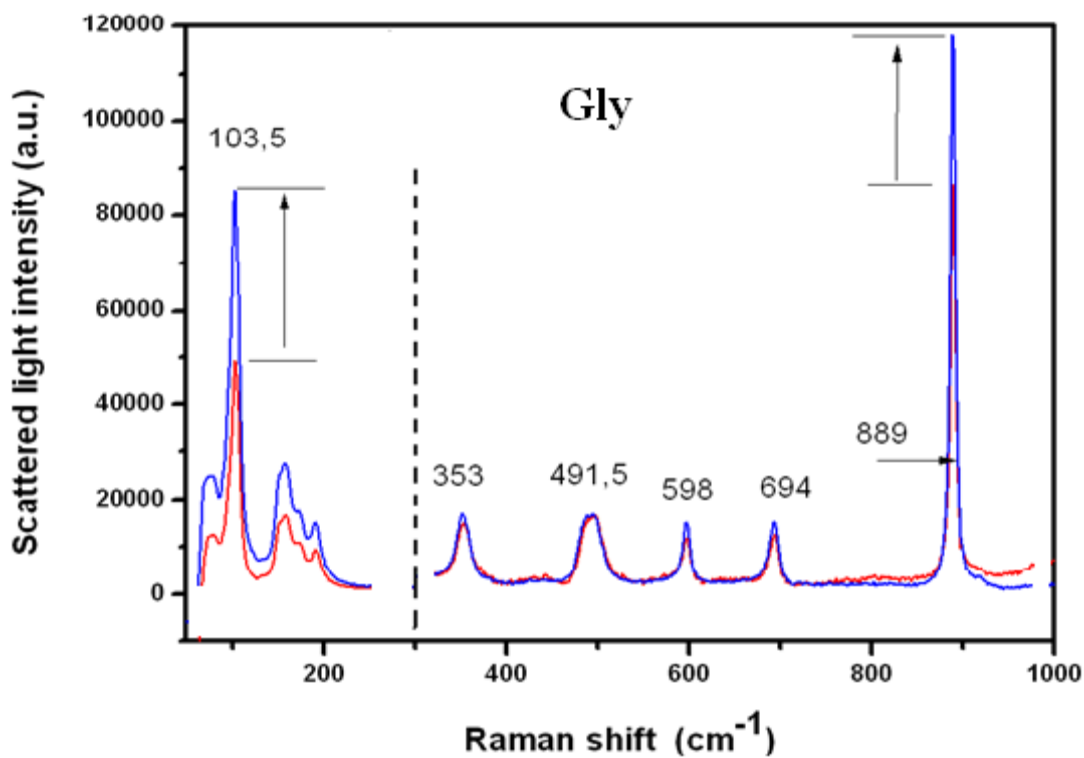

*Рис.4 Спектры рассеяния микрокристаллами глицина в области 70-1000 см$^{-1}$. Интенсивность линий в левой части рисунка уменьшена в 4 раза*

На следующем рисунке 5а,в. представлены спектры рассеяния микрокристаллами триптофана. На верхнем рисунке приводится сравнение исходного и облученного кристалла. На нижнем - сравнение спектра необлученного кристалла и спектра рассеяния от микрокристалла триптофана, полученного из деградировавшего раствора (пожелтевшего от времени) без дополнительно воздействия.

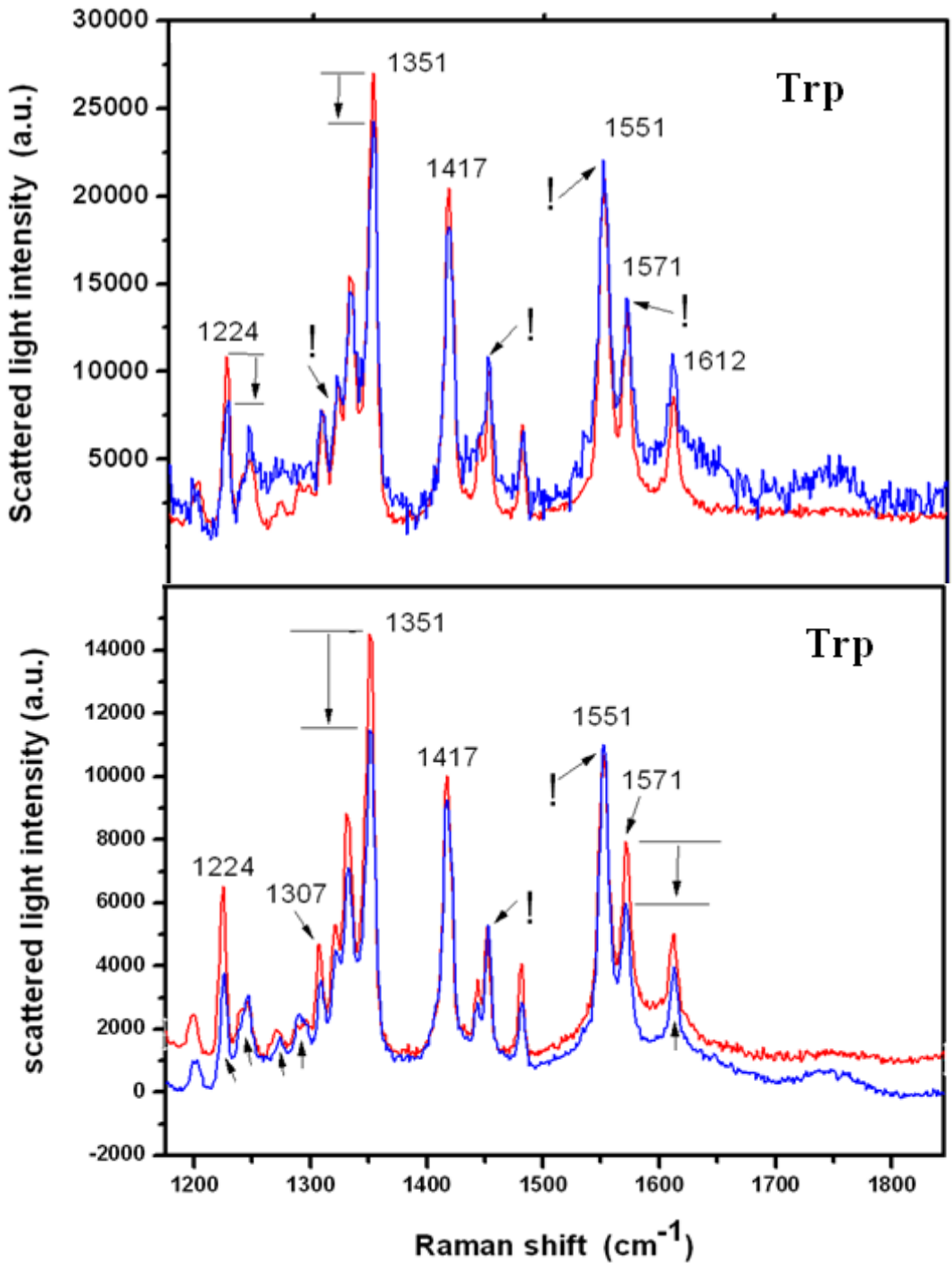

*Рис.5 а,в Спектры рамановского рассеяния микрокристаллами триптофана. Рис.а (верхний) сравнение спектра исходного кристалла, и спектра, полученного после облучения. Рис.в – сравнение спектров микрокристалла исходного материала и микрокристалла триптофана из естественно состаренного раствора. Восклицательные знаки поставлены у линий, для которых относительная интенсивность не изменилась.*

Видно, что относительная интенсивность основной части линий на рис.5 после облучения уменьшилась, в то время как относительные интенсивности линий 1551, 1571 см$^{-1}$ и некоторых других, выделенных на рисунке восклицательными знаками, остались без изменения. На рис.5в спектр того же исходного кристалла сравнивается со спектром монокристалла, полученного высушиванием пожелтевшего после длительного хранения раствора триптофана. Видно, что изменения относительных интенсивностей линий в этих двух случаях подобны.

**Обсуждение результатов.**

Изменение рамановского спектра под действием UV может быть вызвано рядом причин. В работе [5] в качестве возможных механизмов рассматриваются возможность фотоиндуцированных химических реакций. В случае фотолиза с разделением молекул на более мелкие фрагменты можно ожидать исчезновения в спектрах линий, ответственных за определенный тип связей. В более сложных случаях, как например в [7], изменение спектра тканей *in vivo* может свидетельствовать о синтезе других молекул. При этом необходимо учитывать, что спектры могут зависеть и от изменения в окружении исследуемых молекул.

В целом из результатов работы следует отметить устойчивость микрокристаллов исследованных аминокислот к действию ультрафиолетового излучения в относительно малых дозах.

Использованная длина волны ультрафиолетового излучения попадает в полосу поглощения большинства органических молекул, так что возбужденные состояния под действием облучения должны были образовываться. Основной эффект – изменение относительной интенсивности линий, может быть связан с переориентаций молекул в кристаллах. Такая переориентация может идти по разным механизмам. Поглотив квант света и затем возвращаясь в основное состояние, молекула может оказаться в иной ориентации, чему может способствовать существование нескольких модификаций кристаллических структур для одного типа аминокислоты.

Насколько можно судить по данным экспериментов, разложения молекул и образования посторонних веществ не обнаружено. Доводом к этому является неизменный набор компонент спектра с почти не изменяющимися положениями.

Слабые изменения положений, наблюдающиеся в некоторых случаях, скорее всего, связаны с неоднородным уширением линий и преимущественной переориентацией молекул на определенных позициях.

Возможно, слабо структурированный фон под линиями на рис.5а в спектре облученного микрокристалла триптофана свидетельствует о начале процесса разложения материала.

## Литература к статье

**.1.** J.T. Wickramamasinghe, C. Wickramamasinghe, W.M.Napier "Comets and the origin of life" // World Scientific , 2010, ISBN-13 978-981-256-635-5, 232 p.

**.2.** Horneck, Gerda; Klaus, David M.; Mancinelli, Rocco L. (2010). "Space Microbiology". Microbiology and Molecular Biology Reviews. **74** (1): 121–156. doi:10.1128/MMBR.00016-09.

**.3.** А.И.Опарин «Происхождение жизни» // 1924 ((1959 М. Воениздат, 38 стр))

**.4.** A.Jirasek, H.G.Schulze, C.H.Hughesman, A,L.Cgreagh, C.A.Haynes, M.W.Blades, R.F.B.Turner "Discrimination between UVradiation-induced and termallyinduced spectral changes in AT-paired DNA oligomers using resonance raman spectroscopy" // J. RAMAN SPECTROSCOPY 2006, 37, 1368-1380/ DOI: 10.1002./jrs 1552

**.5.** V.G.Borio, A.U.Fernandes, L.Silveira Jr. (2015) "Characterization of an ultraviolet irradiation chamber to monitor molecular photodegradation by Raman spectreoscopy" // Instrumentation science and technology DOI:101080/10739149.2015.1081936

.**6**. H.Jia, G.Chen, W.Wang "UV radiation-induced Raman spectra changes in lead silicate glasses" // Optical Materials 2006, 29, 445-448. DOI:10.1016/j.optmat.2005.09,078

.**7**. M.G.Tosato, D.E.Orallo, S.M.Ali, M.S.Churio, A.A.Martin, L. Dicelio "Confocal raman spectroscopy In vivo biochemical changes in the human skin by topical formulation under UV Radiation" // Photochemistry and Photobiology (B), 2015, 8, 30 DOI:10.1016/j.jphotobiol/2015.08030